# Millimeter Wave Thin-Film Bulk Acoustic Resonator in Sputtered Scandium Aluminum Nitride

Sinwoo Cho, Omar Barrera, Pietro Simeoni, Emily N. Marshall, Jack Kramer, Keisuke Motoki, Tzu-Hsuan Hsu, Vakhtang Chulukhadze, Matteo Rinaldi, W. Alan Doolittle, and Ruochen Lu

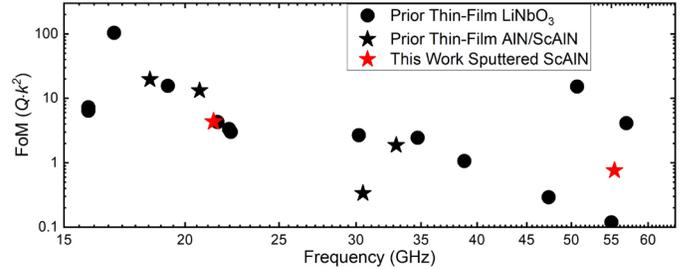

Fig. 1 Survey of reported resonators above 15 GHz.

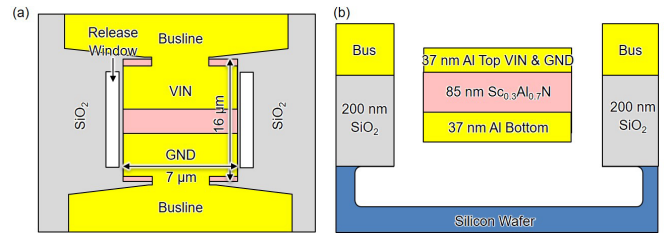

Fig. 2 (a) Top and (b) cross-sectional view of ScAlN mmWave FBAR.

*Abstract*—This work reports a millimeter wave (mmWave) thin-film bulk acoustic resonator (FBAR) in sputtered scandium aluminum nitride (ScAlN). This paper identifies challenges of frequency scaling sputtered ScAlN into mmWave and proposes a stack and new fabrication procedure with a sputtered $Sc_{0.3}Al_{0.7}N$ on Al on Si carrier wafer. The resonator achieves electromechanical coupling ($k^2$) of 7.0% and quality factor ($Q$) of 62 for the first-order symmetric (S1) mode at 21.4 GHz, along with $k^2$ of 4.0% and $Q$ of 19 for the third-order symmetric (S3) mode at 55.4 GHz, showing higher figures of merit (FoM, $k^2 \cdot Q$) than reported AlN/ScAlN-based mmWave acoustic resonators. The ScAlN quality is identified by transmission electron microscopy (TEM) and X-ray diffraction (XRD), identifying the bottlenecks in the existing piezoelectric-metal stack. Further improvement of ScAlN/AlN-based mmWave acoustic resonators calls for better crystalline quality from improved thin-film deposition methods.

*Index Terms*—acoustic resonators, piezoelectric devices, scandium aluminum nitride (ScAlN), millimeter-wave devices, thin-film bulk acoustic resonator (FBAR), thin-film devices

## I. Introduction

Radio frequency (RF) acoustic devices are widely used as sub-6 GHz front-end filters [1]–[4]. Acoustic resonators, i.e., key building blocks for filters, piezoelectrically convert the electromagnetic (EM) energy to mechanical vibrations and efficiently store energy at resonances. Such transduction offers two key advantages over EM counterparts, namely, miniature footprints and better frequency selectivity [2].

Among different piezoelectric RF acoustic platforms, current commercial FBARs have been dominated by sputtered aluminum nitride (AlN) and scandium aluminum nitride (ScAlN) since they possess good acoustic properties high-quality factor ($Q$), and electromechanical coupling ($k^2$) and their well-established microfabrication process that can be integrated into semiconductor industries effortlessly [2], [5].

With the development of wireless communication into millimeter wave (mmWave, >30 GHz) bands, it would be great to frequency scale ScAlN/AlN devices while maintaining high performance for future RF front ends [3]. However, prior studies show that frequency scaling ScAlN/AlN FBARs is challenging (Fig. 1), marked by degraded figures of merit (FoM, $k^2 \cdot Q$) at higher frequencies [6]–[9]. First, it is non-trivial to deposit a high-quality piezoelectric and metal stack, as the required thickness is sub-100 nm for mmWave operation, imposing challenges for conventional sputtering techniques while causing excessive acoustic loss from the thin-film structure [10]. Second, as the resonator's lateral dimensions significantly scale down for 50 Ω systems, acoustic designs and microfabrication procedures for miniature devices are not well studied. More recently, FBARs using better ScAlN/AlN films synthesized by metal-organic vapor phase epitaxy (MOVPE) and molecular beam epitaxy (MBE) have been reported, but the fundamental thickness extensional FBARs show results comparable to their sputtered counterparts at mmWave (Fig. 1) [7], [8], [11], [12]. It is unclear whether the film quality or design/fabrication is the bottleneck for current ScAlN/AlN mmWave FBARs.

In this work, we report a mmWave FBAR using sputtered $Sc_{0.3}Al_{0.7}N$. The resonator achieves $k^2$ of 7.0% and $Q$ of 62 for the first-order symmetric (S1) mode at 21.4 GHz, along with $k^2$ of 4.0% and $Q$ of 19 for the third-order symmetric (S3) mode at 55.4 GHz, showing a higher FoM than reported AlN/ScAlN-based mmWave acoustic resonators. The results are enabled by both the acoustic design and a new fabrication procedure. Material-level analysis indicates that the bottleneck for further performance enhancement calls for better crystalline quality from improved thin-film deposition methods.

Manuscript received 1 August 2023; revised XX August 2023; accepted XX August 2023. This work was supported by DARPA COmpact Front-end Filters at the ElEment-level (COFFEE).
  S. Cho, O. Barrera, J. Kramer, T.-H. Hsu, V. Chulukhadze, and R. Lu are with The University of Texas at Austin, Austin, TX, USA (email: sinwoo@utexas.edu). E. N. Marshall, K. Motoki, and W. A. Doolittle are with Georgia Institute of Technology, Atlanta, GA, USA. P. Simeoni, M. Rinaldi are with Northeastern University, Boston, MA, USA.

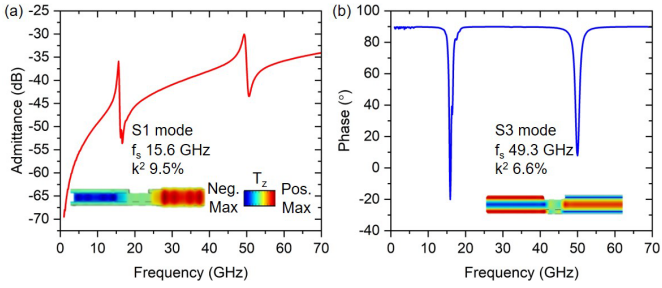

Fig. 3 FEA simulated wideband admittance (a) amplitude and (b) phase. S1 and S3 thickness extension shape modes are plotted.

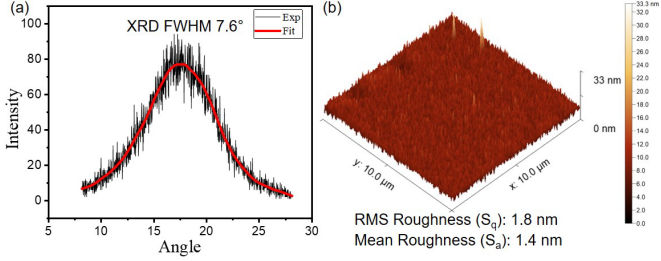

Fig. 4 (a) XRD symmetric rocking curve of 85 nm sputtered thin-film ScAlN. (b) ScAlN film surface roughness measurement by AFM.

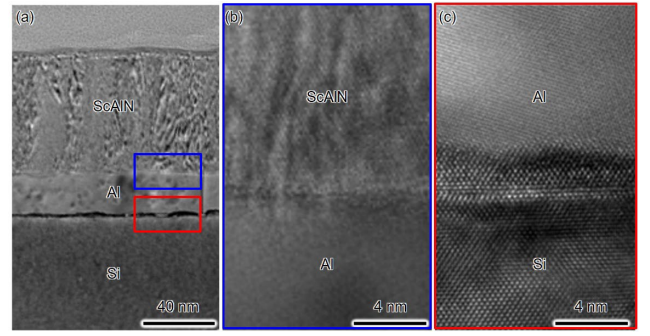

Fig. 5 (a) Cross-sectional TEM images and magnified views of the (b) ScAlN-Al and (c) Al-Si interfaces.

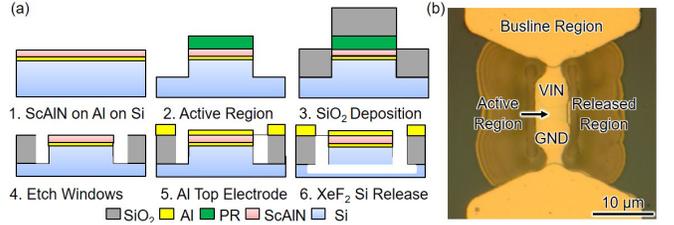

Fig. 6 (a) Device fabrication process and (b) microscopic image of the FBAR.

## II. DESIGN AND SIMULATION

The FBAR top and cross-sectional views are shown in Fig. 2 (a)-(b). The film stack consists of 85 nm thick $Sc_{0.3}Al_{0.7}N$ sandwiched between 37 nm thick aluminum (Al) top and bottom electrodes, with signal and ground traces on the top along with a floating bottom electrode. Such thickness is selected to enable S3 mode around 50 GHz. Due to the high capacitance density of the thin film, the lateral dimensions of the resonant body are designed as 7 μm by 16 μm. The buslines and probing pads are thickened to 300 nm for less routing resistance. A key differentiator for this work is that we start with uniformly sputtering ScAlN on sputtered Al on a silicon (Si) carrier wafer before passivating the majority of the substrate with silicon dioxide ($SiO_2$) except for the active region. First, this allows sputtered films with better quality than those on patterned bottom electrodes, especially near the edge of the patterned bottom electrodes. Second, the $SiO_2$ reduces the feedthrough-induced parasitic capacitance and resistance, which is more pronounced at mmWave [13]. The fabrication process for such structures will be explained in Section III.

The proposed FBAR is simulated [Fig. 3 (a)-(b)] using COMSOL finite element analysis (FEA) with a mechanical $Q$ value of 50, estimated from earlier mmWave AlN FBARs [14]. In operation, the electric field between the top and bottom electrodes excites first-order symmetric (S1) and third-order symmetric (S3) modes via piezoelectric coefficient $e_{33}$. S1 at 15.6 GHz shows $k^2$ of 9.5%, while S3 at 49.3 GHz shows $k^2$ of 6.6%. $k^2$ follows the equation definition in Fig. 7 (e) [15]. The mode shapes are plotted in Fig. 3, confirming that the stack selection maximizes $k^2$ for S3, as the stress nodes lie in the Al-ScAlN interfaces.

## III. MATERIAL ANALYSIS AND FABRICATION

The fabrication starts with sputtering 37 nm of Al and 85 nm of ScAlN onto a high-resistivity ( > 10,000 Ω·cm) Si <100> wafer with an Evatec Clusterline 200 sputtering tool without breaking vacuum. The quantitative material analysis starts with X-ray diffraction (XRD) in Fig 4 (a). The full width at half maximum (FWHM) of the rocking curve is 7.6°, indicating that the sputtered thin film has non-ideal crystal quality, given that it is sputtered on top of metal, while the overall thickness is sub-100 nm. Fig. 4 (b) shows the atomic force microscopy (AFM) and the surface roughness of the sputtered ScAlN film. The surface is generally flat, with a few spikes caused by the defects formed during Al deposition. The film quality of the stack is validated using transmission electron microscopy (TEM) images shown in Fig. 5. The crystal shows misorientation angles as large as 18° [Fig. 5 (a)], which is likely caused by the deformation of the Al layer [Fig. 5 (c)] during sputtering (350 °C process) and could be overcome in the future with platinum (Pt) electrodes upon further development. Such moderate film quality is a bottleneck for future mmWave ScAlN FBARs, calling for better deposition methods.

The fabrication process is shown in Fig. 6 (a). First, the regions outside the active areas composed of ScAlN, Al, and Si layers are etched by AJA Ion Mill. The etched regions are then passivated with a low-temperature (100 °C) plasma-enhanced chemical vapor deposition (PECVD) deposition of 200 nm of $SiO_2$, providing electrical isolation while preventing top electrode disconnection due to the height steps. Next, release windows are defined and etched using the AJA Ion Mill. The top 37 nm Al electrodes and 300 nm thickened Al buslines are then deposited using a KJL e-beam evaporator. Structural release with a xenon difluoride ($XeF_2$) Si isotropic etch is applied to ensure energy confinement within the FBAR. The optical image of the fabricated FBAR is shown in Fig. 6 (b).

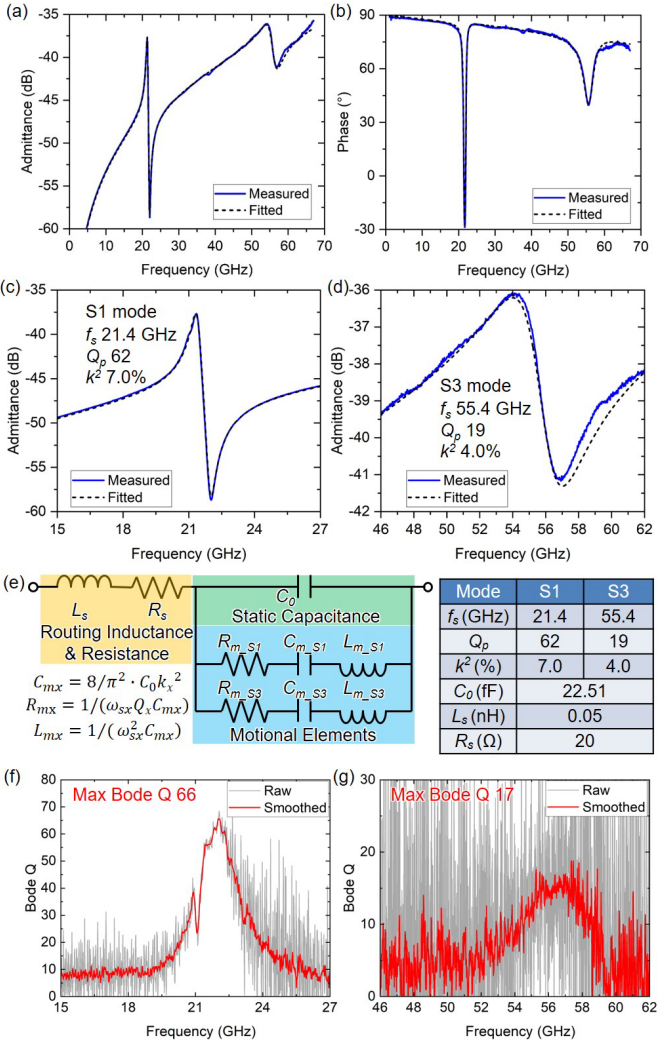

Fig. 7 Measured wideband admittance in (a) amplitude and (b) phase. Magnified admittance for (c) S1 mode at 21.4 GHz and (d) S3 mode at 55.4 GHz. (e) Modified mmWave mBVD model and extracted key resonator parameters. Bode Q for (f) S1 and (g) S3 tones.

## IV. Measurement and Discussion

The resonator is measured using a Keysight vector network analyzer (VNA) in room temperature air at −15 dBm power level. Two-port measurement is performed [16]. The measured admittance amplitude and phase are plotted in Fig. 7 (a)-(b), showing S1 at 21.4 GHz and S3 at 55.4 GHz. The minor resonance between S1 and S3 is the second-order antisymmetric (A2) mode due to the slight thickness difference in the top and bottom electrodes. The admittance curves from Fig. 7 (a)-(b) are magnified and plotted in Fig. 7 (c) for S1 and Fig. 7 (d) for S3. To extract the resonator performance, a modified mmWave modified Butterworth Van Dyke (mBVD) model is used [Fig. 7 (e)], adding series routing resistance ($R_s$) and inductance ($L_s$) for capturing the EM effects. The EM parameters, i.e., $R_s$, $L_s$, and static capacitance $C_0$, are first fitted from the admittance amplitude and phase [Fig. 7 (a)-(b)], before adding in the motional elements for extracting $Q$ and $k^2$ in Fig. 7 (e). The fitted curves are plotted in Fig. 7 (a)-(d). The extracted parameters are listed in Fig. 7.

The resonator achieves $k^2$ of 7.0% and $Q$ of 62 for S1, along with $k^2$ of 4.0% and $Q$ of 19 for S3, leading to a FoM of 4.34 and 0.76 respectively. $k^2$ is extracted via fitting in Fig. 7. Note that $Q$ here is effectively the anti-resonance quality factor $Q_p$. To further validate, Fig. 7 (f)-(g) display the Bode $Q$ [17] for S1 and S3, respectively. The maximum Bode $Q$ after smoothing is 66 for S1 and 17 for S3.

To compare with the state of the art (SoA), a survey of FoM for reported resonators above 15 GHz is reported in Fig. 1, including both the AlN/ScAlN [6]–[9] and lithium niobate (LiNbO$_3$) demonstrations [18]–[22]. Despite moderate film quality, our work shows comparable FoM to earlier ScAlN/AlN work for the 21.4 GHz, while the 55.4 GHz devices show higher FoM than earlier ScAlN/AlN works, proving the effectiveness of the new stack. However, the FoM is lower than that of transferred thin-film LiNbO$_3$ based mmWave resonators with much better film quality (<100 arcsec FWHM) [16], implying that further improvement of ScAlN/AlN-based mmWave resonators requires better crystalline quality from improved thin-film deposition methods.

## V. Conclusion

We demonstrate a mmWave ScAlN FBAR operating at 21.4 GHz (S1 mode) and 55.4 GHz (S3 mode). The resonator achieves $k^2$ of 7.0% and $Q$ of 62 for S1 at 21.4 GHz, along with $k^2$ of 4.0% and $Q$ of 19 for S3 at 55.4 GHz, showing higher FoM than reported AlN/ScAlN-based mmWave acoustic resonators. Material-level analysis and device-level performance indicate that the bottleneck for further performance enhancement lies in better thin-film deposition methods.

## Reference

[1] S. Gong et al., *IEEE Journal of Microwaves*, vol. 1, no. 2, pp. 601–609, 2021.
[2] R. Ruby, *IEEE Microw Mag*, vol. 16, no. 7, pp. 46–59, 2015.
[3] A. Hagelauer et al., *IEEE Journal of Microwaves*, 2022.
[4] R. Lu et al., *J. Micromech. Microeng.*, vol. 31, no. 11, p. 114001, 2021.
[5] G. Piazza et al., *J Microelectromech Syst*, vol. 15, no. 6, pp. 1406–1418, 2006.
[6] Izhar et al., *IEEE Electron Device Letters*, 2023,
[7] R. Vetury et al., *2023 IEEE/MTT-S International Microwave Symposium - IMS 2023*, pp. 891–894, Jun. 2023,
[8] S. Nam et al., *IEEE Microwave and Wireless Technology Letters*, vol. 33, no. 6, 2023,
[9] Z. Schaffer et al., *IEEE Microwave and Wireless Components Letters*, vol. 32, no. 6, 2022,
[10] S. Cho et al., *IFCS-EFTF 2023 - Joint Conference of the IEEE International Frequency Control Symposium*, 2023.
[11] W. Zhao et al., *J Appl Phys*, vol. 132, no. 2, 2022,
[12] M. Park et al., *J Microelectromech Syst*, vol. 29, no. 4, 2020,
[13] O. Barrera et al., *arXiv preprint arXiv:2307.04559*, 2023.
[14] Z. Schaffer et al., *IEEE Microwave and Wireless Components Letters*, vol. 32, no. 6, pp. 656–659, 2022,
[15] R. Lu et al., *J Microelectromech Syst*, vol. 28, no. 2, pp. 209–218, 2019.
[16] J. Kramer et al., *arXiv preprint arXiv:2307.05742*, 2023.
[17] D. A. Feld et al., *IEEE Int. Ultrason. Symp.*, pp. 431–436, 2008.
[18] Y. Yang et al., *IEEE Trans Microw Theory Tech*, vol. 68, no. 12, 2020.
[19] M. Hara et al., *Proc IEEE Ultrason Symp*, pp. 851–854, 2009,
[20] R. Lu et al., *J Microelectromech Syst*, vol. 29, no. 5, 2020.
[21] S. Link et al., *IEEE International Ultrasonics Symposium, IUS*, 2021,
[22] R. Lu et al., *J Microelectromech Syst*, 2020.